\pdfoutput=1

\documentclass[11pt]{article}
\usepackage{jheppub}

\usepackage{caption}
\usepackage{subcaption}
\usepackage{extarrows}

\usepackage{mathrsfs}

\usepackage{amsmath}
\usepackage{latexsym,amssymb}
\usepackage{graphicx,color,slashed}
\usepackage{bbm} 
\usepackage{mathtools}
\usepackage{ifpdf}

\usepackage{esint}

\usepackage{amsfonts}
\usepackage{amsmath}
\usepackage{epsfig}
\usepackage{isomath}
\usepackage{latexsym,amssymb}
  \usepackage{amsmath,amssymb,amsthm}
  \DeclareSymbolFontAlphabet{\Scr}{rsfs}

\newcommand{\cL}{\mathcal{L}}
\newcommand{\cM}{\mathcal{M}}
\newcommand{\cN}{\mathcal{N}}

\newcommand{\be}{\begin{equation}}
\newcommand{\ee}{\end{equation}}
\newcommand{\ba}{\begin{eqnarray}}
\newcommand{\ea}{\end{eqnarray}}

\def\E{{$E_{7(7)}$}}

\def\ID{\relax{\rm I\kern-.18em D}}

\newcommand{\rf}[1]{(\ref{#1})}
\newcommand{\bea}{\begin{eqnarray}}
\newcommand{\eea}{\end{eqnarray}}

\def\bfzero{\relax{\rm I\kern-.18em 0}}
\def\bfone{\relax{\rm 1\kern-.35em 1}}
\def\twomat#1#2#3#4{\left(\begin{array}{cc}
\end{array}
\right)}

\newcommand{\C}{\mathds{C}}

\title{\rm{\bf   Enhanced  Duality  in 4D Supergravity}}
\author
{ Renata Kallosh}
 \affiliation{Stanford Institute for Theoretical Physics and Department of Physics,\\
 Stanford University, Stanford, CA 94305, USA\\
}

\abstract{
 U-duality imposes strong constraints on the structure of divergences in supergravity. But Gaillard-Zumino  Sp$(2n_v, \mathbb{R})$ duality in D=4 has more symmetries than U-duality.  For example, in $\cN = 8$ the dimension of Sp(56) is 1596, whereas its U-duality subgroup  \E\ has a dimension of 133. In comparison, in  D $> 4$ maximal dualities are U-dualities. We argue that the extra dualities in 4D, enhancing U-duality, determine the properties of perturbative quantum supergravity.  The presence/absence of enhanced dualities suggests a possible explanation of the results of the amplitude loop computations in D-dimensional supergravities and of the special status of D $= 4$ in this respect.
}

\begin{document}

\maketitle

%\tableofcontents{}

%\newpage

\parskip 5pt

%%%%%%%%%%%%%%%%%%%%%%%%%%%%%%%%%%%%%%%%%%%%%%%%%%%%%%%%%%%%%%%%%%%%%%

\section{Introduction}
In the recent papers  \cite{Kallosh:2023css}  we analyzed
various computations performed in diverse dimensions in perturbative supergravity \cite{Salam:1989ihk,Bern:2023zkg}. A simple observation made in  \cite{Kallosh:2023css} is that {\it all} $D=5,6,7,8,9,10,11$ maximal supergravities have UV divergences at some loop order. But in  $D = 4$, all presently available computations for supergravities $\cN\geq 5$ did not detect any UV divergences.

One may wonder whether there is any simple reason why the results of calculations for $\cN\geq 5$ in $D= 4$ are different from those in $D> 4$.  We believe that such a reason is that {\it only in $D=4$ dualities are enhanced}. 
U-duality symmetries in maximal supergravities are known to form a chain, $E_{11-D(11-D)}$,  when going up from 3D all the way to 11D. This process was described by B. Julia \cite{Julia} as a Group Disintegration
\be
E_{8(8)} \supset E_{7(7)} \supset E_{6(6)}  \supset E_{5(5)}  \supset E_{4(4)}  \supset E_{3(3)}  \supset E_{2(2)} \supset E_{1(1)}  \supset E_{0(0)}  \ .
\ee
Hull and Townsend \cite{Hull:1994ys} studied duality symmetries in supergravity and string theory and called these symmetries U-dualities, these are continuous in supergravity and discrete in string theory. In supergravities, equations of motion are U-duality invariant, and the relevant groups are G$_U= E_{d+1(d+1)}$ with $d=10-D$.  

It is also known  that  Gaillard-Zumino (GZ) electro-magnetic duality symmetry \cite{Gaillard:1981rj}
in 4D supergravity is bigger than U-duality. GZ group is  $Sp(2 n_v, \mathbb{R})$, where $2n_v$ is a number of vectors in electro-magnetic duality doublet.  These are 56, 32, 20 for $\cN=8, 6, 5$, respectively.  Dimension of $Sp(2 n_v, \mathbb{R})$ exceeds significantly the dimension of the U-duality  groups G$_U$:  $E_{7(7)}, SO^*(12), SU(1,5)$ (known as groups of type E7) for $\cN=8, 6, 5$, respectively. In all odd dimensions, GZ duality is absent; in even dimensions $D>4$, one can check case by case that maximal duality coincides with $G_U$ duality. Therefore
\bea \label{max4D}
&&D =4 :\qquad \rm{dim } \, [{\rm maximal \, duality } ]=\rm{dim } [ Sp(2 n_v, \mathbb{R})] \gg \rm{dim } [G_U] \ ,\\
&& D>4  :\qquad \rm{dim } \, [{\rm maximal \, duality } ] = \rm{dim } [G_U] \ . 
\label{maxD}\eea
Extended supergravities have physical scalars in  ${G\over H}$ coset space, where G is the U-duality group, and H is its maximal compact subgroup. 
 GZ duality gives a significant enhancement to U-duality in 4D. Its role became more clear after the D-dimensional computational data \cite{Bern:2023zkg} were analyzed in \cite{Kallosh:2023css} in the context of local H symmetry and global G symmetry \cite{WIP1, WIP2}.
 
It is not widely recognized that there are two different types of supergravities in dimension D,  with the same amount of local supersymmetry. Here and in  \cite{WIP1, WIP2}, we call ``supergravities of type I'' the well-known supergravities with local  $H_D$ symmetry and global $E_{D-11(D-11)}$ U-duality. We call ``supergravities of type II'' the ones derived from higher dimensions without dualization. Their manifest local and global symmetry groups are smaller,  inherited from higher dimensions.

In both types of supergravities, one can gauge-fix their local symmetries. The existence of these different supergravities and the ability to perform the gauge-fixing in various gauges, symmetric or Iwasawa-type gauges, raise the question about the quantum equivalence of these different theories. Note that the loop computations of S-matrix in amplitudes do not differentiate between these.

 In the past only the \E\, subgroup of $Sp(56, \mathbb{R})$ was considered in the investigation of candidate superinvariant counterterms in  \cite{Kallosh:2011dp}: it was based on the on-shell superspace of maximal supergravity in \cite{Brink:1979nt}, where only \E\, subgroup of $Sp(56, \mathbb{R})$ duality is present in superspace with local  SU(8) H-symmetry. 

Now, however, we believe that the full $Sp(2 n_v, \mathbb{R})$ symmetry plays a significant role in quantum supergravity. It relates various versions of supergravities gauge-fixed in different gauges. This conclusion is based on studies of {\it different symplectic frames} in 4D supergravity related to each other by an $Sp(2 n_v, \mathbb{R})$ transformation  \cite{deWit:2002vt} combined with the {\it Hamiltonian formulation of dualities} \cite{Henneaux:2017kbx}. We have found that the $Sp(2 n_v, \mathbb{R})$ symmetry is a bona fide off-shell symmetry of the Hamiltonian path integral defining perturbative quantum theory.

In this paper, we will argue that the results obtained in \cite{deWit:2002vt,Henneaux:2017kbx} in combination with the analysis of amplitude loop 
computations performed in \cite{Kallosh:2023css} suggest  that there is an enhancement of dualities in 4D affecting quantum perturbative $\cN\geq 5$ supergravity.\footnote{
 Half-maximal supergravities require special treatment  \cite{Kallosh:2023css}, \cite{WIP2}.}

\section{UV divergences and broken local H and global G symmetries}

An on-shell superspace construction of D-dimensional  G/H supergravity is based on local H-symmetry and global G-symmetry. Local Lorentz symmetry, together with local H-symmetry, form the set of symmetries of the tangent space of the on-shell superspace. In $D = 4$, detailed constructions are available \cite{Brink:1979nt}; in higher $D$,  construction is analogous.

The geometric on-shell superinvariants are available starting from the loop order $L_{cr}$. We have defined $L_{cr}$   as the loop order where local H-symmetry and global G-symmetry of the candidate counterterms are available
\cite{Kallosh:2023css}. It was established there that for $\cN\geq 5 $ D-dimensional supergravity
superinvariants with local H-symmetry and global G-symmetry  are available starting with 
\be
 L_{cr}= {2\cN +n \over (D-2)} \ , \qquad n\geq 0 \ .
\label{cr}\ee
In particular, in $D=4$,  $L_{cr}= \cN $  and  $L_{cr}=8, 6, 5$ for $\cN=8,6,5$ respectively \cite{Kallosh:1980fi}.
In each case in \rf{cr}, we need to find a minimal value of $n$, which makes $ L_{cr}$ an integer\footnote{Supergraity actions with specific global G and local H symmetries are available only in integer dimensions.}. The result for $ L_{cr}$ for the case of maximal supergravity is shown below. We also show the computational data in D-dimensional supergravities \cite{Bern:1998ug},  at which   loop order  in each dimension  UV divergence was detected
\bea 
 &D=4, \, L_{cr}=8 :  \quad &\kappa^{14} \int d^4\, x \, D^{10} R^4 +\dots \,     \quad n=0  \cr
&D=5,  \, L_{cr}=6 :  \quad &\kappa^{10} \int d^5\, x \, D^{12} R^4+\dots \,        \quad n=2    \qquad \qquad L_{UV}=5 < L_{cr}=6\cr
& D=6,  \, L_{cr}=4 :  \quad &\kappa^{6} \int d^6\, x \, D^{10} R^4+\dots \,    \, \,    \quad n=0  \qquad \qquad L_{UV}=3 <L_{cr}=4\cr
& D=7,  \, L_{cr}=4 :  \quad &\kappa^{6} \int d^7\, x \, D^{14} R^4+\dots \,    \,     \quad \, n=4  \qquad \qquad L_{UV}= 2 < L_{cr}=4\cr
& D=8,  \, L_{cr}=3 :  \quad &\kappa^{4} \int d^8\, x \, D^{12} R^4+\dots \,    \,     \quad \, n=3  \qquad \qquad L_{UV} = 1 <  L_{cr}=3 \cr
& D=9,  \, L_{cr}=3 :  \quad &\kappa^{4} \int d^9\, x \, D^{15} R^4+\dots \,    \,     \quad \, n=5 \qquad \qquad  L_{UV} = 2 <  L_{cr}=3 
\label{Table}\eea
 The data has revealed a universal feature for all maximal $D>4$ supergravities: UV divergences appear at the loop order {\it below critical}.
Thus, all UV divergences in maximal $D>4$ supergravities imply that the local H-symmetry and global G-symmetry are broken: the superinvariants supporting these UV divergences are given by the superspace integrals with the volume of integration, which is a subspace of the total supervolume \cite{Kallosh:2023css}. 

 The loop computations tell us that so far, no UV divergences in 4D at $\cN\geq 5$ below $L_{cr}$ were found. For example, in $\cN=5$ there is no UV divergence at $L=4 < L_{cr}= 5$ \cite{Bern:2014sna}. 
There is a striking difference between the D=4 and D$>$4 in the dimension of maximal dualities versus U-dualities in eqs. \rf{max4D} and  \rf{maxD}. The same difference we see   in eq. \rf{Table}  presenting  UV divergences discovered at present.

Is it possible that these extra symmetries beyond U-dualities, absent in \rf{maxD}  and 
present in \rf{max4D},  can explain the local H-symmetry and global G-symmetry anomalies in D$>$4 associated with UV divergences, and the absence of these in 4D $\cN\geq 5$ so far?
As we will argue below,  the answer to this question is positive. A detailed investigation of this issue is performed in  \cite{WIP1, WIP2}.  

\section{ A tale of two supergravities in dimension D} 

The role of extra duality symmetries present in 4D and absent in D$>$4 is to relate to each other 4D supergravities quantized in different gauges of local H symmetry, as well as to relate supergravities I and II, as described in \cite{deWit:2002vt} and \cite{WIP1, WIP2}.   We study mostly the case of D-dimensional supergravities II derived from D+1 dimension. 
In case I it is possible to perform the gauge-fixing of a local $H_D$-symmetry in the symmetric gauge where all scalars enter the action non-polynomially. In case II, there is a positive minimal number of axions,  which enter the action polynomially. This number is 
defined by an abelian nilpotent ideal of the solvable Lie algebra \cite{Andrianopoli:1996zg}.

The role of the gauge fields of local H-symmetry is played by the composite scalar-dependent connections. Theory I has local $H_D$-symmetry, and Theory II has a local $H_{D+1}$-symmetry,  which can be gauge-fixed in various gauges.
The unitary gauges in extended supergravities are studied in \cite{WIP1}, these include symmetric and Iwasawa-type gauges. Symmetric gauges tend to preserve global H symmetry and nonlinearly realized G symmetry on-shell. Iwasawa-type gauges of D-dimensional supergravities studied in \cite{WIP1} have, in general,  less symmetries since they are inherited from (D+1)-dimensional supergravities compactified on a circle.

For example, in 4D  there is a maximal supergravity  I  \cite{Cremmer:1979up} with 70 scalars in ${E_{7(7)}\over SU(8)}$
coset space which has  global on-shell $SU(8)$ symmetry in the symmetric gauge 
$
\phi_{ijkl} = {1\over 4!} \epsilon_{ijklpqrs} \bar \phi^{pqrs}
$.
And there is   supergravity II in 4D derived in \cite{Andrianopoli:2002mf} from a 5D maximal supergravity  where  
${\bf 70}$ SU(8) scalars of  supergravity I   are decomposed  in $E_{6(6)} \times SO(1,1)$  as ${\bf 70}\to \,  \phi  +a^\Lambda  + \phi^{abcd}$
\be
{\bf 70}\to {\bf 1}+{\bf 27}+{\bf 42} \ .
\ee
The 42 scalars are in the coset space ${E_{6(6)}\over USp(8)}$.
Also the  {\bf 56} vectors of SU(8) are decomposed in $E_{6(6)} \times SO(1,1)$.
One can consider supergravity I in a different symplectic frame, not the $SL(8, \mathbb{R})$ frame as in \cite{Cremmer:1979up} but in $E_{6(6)}$ frame as in \cite{deWit:2002vt}. In this new frame, a different choice of the gauge can be taken so that resulting supergravity I has the field structure of 
 supergravity II \footnote{It is amazing  that the non-BPS extremal black holes \cite{Ferrara:2006em} in maximal 4D supergravity  are actually solutions of supergravity II \cite{Andrianopoli:2002mf}, whereas the  1/8 BPS extremal black holes are solutions of supergravity I \cite{Cremmer:1979up}.
}.

In 6D the situation is the following: there is supergravity I \cite{Tanii:1984zk} where scalars are in ${E_{5(5)}\over SO(5)\times SO(5)}$
coset space which has a global $SO(5)\times SO(5)$ symmetry in the symmetric gauge. The 25 scalars are 
$
\phi^{a\dot a} =  \gamma^a_{\alpha \beta } \gamma^{\dot a}_{\dot \alpha \dot \beta} \phi^{\alpha \beta \dot \alpha \dot \beta}
$.
But there is also supergravity II derived from a 7D maximal supergravity \cite{Cowdall:1998rs} where 25 scalars are 
decomposed in $E_{4(4)}\times O(1,1)$ as ${\bf 25}\to   \phi  +B^{[IJ]}  + \Pi_I{}^j$
\be
{\bf 25}\to {\bf 1}+{\bf 10}+{\bf 14} \ .
\ee
Here $\Pi_I{}^j$ is a real symmetric unimodular matrix, a coset representative of ${E_{4(4)}\over SO(5)}$. Analogous decomposition takes place for 2-tensors.

Note that both supergravity I and supergravity II actions in dimension D have the maximal amount of local supersymmetries. The on-shell global symmetry $E_{5(5)}$ is present in supergravity I, but in supergravity II, only $E_{4(4)}$  is manifest, as inherited from D+1.

\section{ Symmetric and Iwasawa-type  gauges  in  supergravity I}
D-dimensional supergravities I, which have local $H_D$-symmetry, can be quantized \cite{WIP1} in symmetric gauge or in Iwasawa-type gauges representing the states inherited from D+1-dimensional supergravity, compactified on a circle, as in supergravities II.  The relation between these gauges was studied in \cite{WIP2}, based on the construction of different symplectic frames in 4D in \cite{deWit:2002vt}.

We are interested in 2 frames, the original one  ${\bf SL(8, \mathbb{R})}$ {\bf frame}, where the action has  manifest  $SL(8, \mathbb{R})$ symmetry, and ${\bf E_{6(6)}}$ {\bf frame}, where the action has 
 manifest $E_{6(6)}\times SO(1,1)$ symmetry inherited from 5D maximal supergravity. 
 
These different symplectic frames can be reached from one to another using $Sp(56,\mathbb{R})$ GZ transformation, which leads to different Lagrangians. Off-shell, these Lagrangians are different, but the theories are classically equivalent on-shell  \cite{Gaillard:1981rj,deWit:2002vt}. 

A caveat here is that not every $Sp(56, \mathbb{R})$  GZ transformation leads to a different Lagrangian, there is a condition. Some of these new Lagrangians, although they might look different from the original ones, can be brought to their original form using the local change of variables, which are integration variables in the path integral.  Therefore, the construction of new symplectic frames in 4D in \cite{deWit:2002vt} requires a GZ transformation modulo local redefinition of the fields in the Lagrangian, scalars, and vectors. The remaining generators of $Sp(2n_v,\mathbb{R})$ GZ transformation form a double quotient
\be
E^{{4D}}=G_U (\mathbb{R})\backslash Sp(2n_v,\mathbb{R})/GL(n_v , \mathbb{R}) \ .
\label{dq}\ee
For example, in maximal 4D supergravity, the relevant quotient is
\be
E_{_{\cN=8}}^{{4D}}=E_{7(7)}(\mathbb{R}) \backslash Sp(56, \mathbb{R}) / GL(28, \mathbb{R}) \ .
\label{N8}\ee
In general, in 4D for $\cN=8,6,5$, the dimension of the double quotient is a positive number. For example for  $\cN=8$  in \rf{N8} we have
\be
{\rm dim} [E_{_{\cN=8}}^{4D}]= 1596-133-784=679 \ .
\ee
The gauge-fixing in symmetric gauge in 4D supergravity takes place starting with the original action with a local SU(8) symmetry in \cite{Cremmer:1979up}. This action has a manifest symmetry $SL(8, \mathbb{R})$, which is a subgroup of \E\ and corresponds to an $SL(8, \mathbb{R})$ frame. There is a 
56-bein ${\cal V}$ interpolating  from \E\, to local SU(8). 
A choice of the ${E_{7(7)}\over SU(8)}$
coset space representative for a vielbein ${\cal V}$ corresponding to a Cartan decomposition of the Lie algebra of \E\, leads to a maximal 4D supergravity in a symmetric gauge.

The gauge-fixing in the Iwasawa gauges in 4D supergravity proceeds as follows: first one transforms to a new symplectic   $ E_{6(6)}$  frame, still preserving local $SU(8)$ symmetry using a construction in \cite{deWit:2002vt}. Now in this Lagrangian there is 56-bein $\hat {\cal V}$ interpolating from $Sp(56)$ to local SU(8). A choice of the 
coset space representatives for a vielbein $\hat {\cal V}$ corresponding to Iwasawa decomposition of the Lie algebra of \E\, leads to a maximal 4D supergravity in  Iwasawa-type gauge. In this gauge supergravity I has the same set of scalars and vectors as supergravity II.

The relation between these gauges is defined by the properties of the GZ duality: the Lagrangian changes from one symplectic frame to another, however, the on-shell physical observables are the same.
This suggests the gauge independence of observables in the symmetric and Iwasawa-type gauges in maximal 4D supergravity as well as in type I and type II supergravities.

In 6D and 8D, there are also supergravities I and supergravities II, and it is possible to choose a symmetric or an Iwasawa-type gauge in supergravity I when gauge-fixing local H-symmetry \cite{WIP1}. For example, in 6D 
in supergravity I \cite{Tanii:1984zk} there is a local $SO(5)\times SO(5)$ symmetry, but only one orthogonal frame is known, the one with $GL(5, \mathbb{R})$ symmetry. The double quotient analog of the one in 4D in eq. \rf{N8} is trivial
\be
E_{_{\cN=8}}^{6D}=E_{5(5)}(\mathbb{R}) \backslash SO(5,5, \mathbb{R}) / G_{v,t}( \mathbb{R}) \ ,
\label{6D}\ee
since $E_{5(5)}\sim SO(5,5)$, and the same for 8D. 

In odd D, there is no GZ duality.
Thus, there is no known way to establish on-shell equivalence between symmetric and Iwasawa-type gauges in maximal supergravities in D$>$4 or between cases I and II: this indicates that the local H  and global G symmetries might be broken. This entails H-symmetry and G-symmetry anomalies and UV divergences, in agreement with the computational data in \cite{Bern:2023zkg}, which we have illustrated in eq. \rf{Table}.

\section{Hamiltonian path integral and quantum equivalence in 4D}
A classical scalar-vector 4D maximal supergravity action is given in the form \cite{deWit:2002vt,Henneaux:2017kbx}
\be
  e^{-1}\cL_{vector} = - \frac{1}{4} \mathcal{I}_{IJ}(\phi) F^I_{\mu\nu} F^{J\mu\nu} + \frac{1}{8} \mathcal{R}_{IJ}(\phi)\, \varepsilon^{\mu\nu\rho\sigma} F^I_{\mu\nu} F^J_{\rho\sigma}  \ .
\label{vec}\ee
 Here  $ F^I_{\mu\nu} = \partial_\mu A^I_\nu -  \partial_\nu A^I_\mu $.
In this action, an arbitrary symplectic frame is possible, as well as any unitary gauge for a local H-symmetry.  $Sp(56, \mathbb{R})$ symmetry acts on a   doublet of vector field strengths $F$ and their magnetic duals $G$ (derivative of the Lagrangian over $F$).
This doublet transforms under $Sp(56, \mathbb{R})$
\be
\left(\begin{array}{c}F \\G\end{array}\right)'= \left(\begin{array}{cc}A & B \\C & D\end{array}\right) \left(\begin{array}{c}F \\G\end{array}\right) \ ,
\label{doublet}\ee
and the scalars $\phi$ transform accordingly. 
It is convenient to define the symplectic scalar dependent metric using the matrices $\mathcal{I}_{IJ}(\phi)$ and  $\mathcal{R}_{IJ}(\phi)$ in the action \rf{vec}
\begin{eqnarray}\label{56bein}
 {\cal M} =\left(
                                        \begin{array}{cc}
                                         \mathcal{I} + \mathcal{R}\mathcal{I}^{-1}\mathcal{R} &\, \, \,  - \mathcal{R} \mathcal{I}^{-1}  \\
     
                                        - \mathcal{I}^{-1} \mathcal{R} &\, \, \,    \mathcal{I}^{-1} \\
                                        \end{array}
                                      \right)        \ .                         
                                      \end{eqnarray}         
It transforms under $Sp(56, \mathbb{R})$  with $E= \left(\begin{array}{cc}A & B \\C & D\end{array}\right) $ as follows
\be
 {\cM}'(\phi)= E \cM(\phi)  E^T \ .
\label{M}\ee   
When the matrix E is lower triangular, {\it i.e.} $B=0$, the Lagrangian  \rf{vec} is invariant, up to a total derivative \cite{Gaillard:1981rj,deWit:2002vt}. This is the case of the electric subgroup of $Sp(56, \mathbb{R})$. It means that the symplectic frame remains the same, but the coset representative (equivalent to a choice of the gauge) has changed. However, when $B\neq 0$, the off-shell Lagrangian is not invariant, and the theory is described by a different symplectic frame \cite{deWit:2002vt}. On-shell, however, equations of motion as well as Bianchi Identities are the same for any choice of the coset representatives and any choice of the symplectic frame. This also covers the cases of supergravity I and supergravity II, which classically on-shell are the same, as shown in \cite{deWit:2002vt}.

To promote this equivalence to a quantum theory, we have to switch from Lorentz covariant to the 1st order formulation since dualities act on $F_{\mu\nu}$'s instead of acting on vector fields, which are integration variables one can change in the path integral.

In the flat space, {\it i.e.} in the absence of gravitational interaction,   bosonic action of supergravity vectors interacting with scalars can be used, following the 1st order formulation in \cite{Henneaux:2017kbx},  to define a path integral for the S-matrix \cite{WIP2}
 \be
\langle {\rm out} | S | {\rm in} \rangle= \int \exp \Big ( {i\over h} \int d^4x  (P_i ^T \dot Q^i- P^T {\cal M}(\phi) P) )\Big) \prod_x \delta(P^i-P^i(Q))dP_i (x)dQ^i (t) \ .
\label{S}\ee
Here both $P$'s and $Q$'s are  $Sp(56, \mathbb{R})$ doublets and space-time 3-vectors, $i=1,2,3$
\be
Q_i^M= \mathcal{A}^M_i \, ,\qquad P^i_M = \Omega_{MN}\mathcal{B}^{N i}\, , \qquad 
\mathcal{B}^{N i}=  \varepsilon^{ijk} \partial_j \mathcal{A}^N_k  \ .
\ee
This Hamiltonian path integral is an example of a general construction for gauge theories by Faddeev and Fradkin-Tyutin-Batalin-Vilkovisky, described in a review by Henneaux \cite{Faddeev:1969su}.

One finds that to change the S-matrix to reach a different symplectic frame or a different gauge of the local H-symmetry, or to reach supergravity II starting from supergravity I corresponds to a canonical change of variables with some specific choices of the $Sp(56, \mathbb{R})$ matrix of the form given in \rf{doublet}. These choices involve an electric subgroup of $Sp(56, \mathbb{R})$ with a vanishing $B$ in eq. \rf{doublet} to change the gauge and a double quotient matrix in eq. \rf{dq} to change the symplectic frame. The relevant transformation of the scalar dependent matrix $\cM$ takes the form given in eq. \rf{M}. The S-matrix is the same for all these cases due to a canonical change of variables in the path integral.

\section{Summary}

Here we described enhanced duality in 4D supergravity due to Gaillard-Zumino $Sp(2n_v, \mathbb{R})$ duality \cite{Gaillard:1981rj},  based on the recent work in \cite{WIP1, WIP2}. The quantitative measure of the enhancement of duality is a dimension of the double quotient introduced in \cite{deWit:2002vt}, see eq. \rf{dq}, \rf{N8} here. 
Combining theory with amplitude data, one can understand the pattern, explain the data, and make predictions.
The important new insights come from combinations of many  ideas: 

\begin{itemize}

\item{Hamiltonian path integral  construction for gauge theories by
Faddeev-Fradkin-Tyutin-Batalin-Vilkovisky-Henneaux  \cite{Faddeev:1969su},  1969}.

  \item 4D Gaillard-Zumino duality \cite{Gaillard:1981rj}, 1981, $Sp(56) \supset $ \E .
  
  \item Andrianopoli, D'Auria, Ferrara, Fr\'e, Minasian, Trigiante \cite{Andrianopoli:1996zg},  1996:  supergravities in dimension D derived from D+1 supergravity.
 Abelian ideal of the solvable Lie algebra  = number of axions in supergravity II and in partial Iwasawa gauges of supergravity I, generic in all dimensions. 
   
  \item DeWit, Samtleben, Trigiante  \cite{deWit:2002vt}, 2003: symplectic frames and double quotients. Duality symmetry with non-trivial double quotients, when available, serves to prove classical equivalence of symmetric and Iwasawa-type gauges in supergravities I and II and, in this way, suggests a possibility of the absence of local H-symmetry anomalies.
  
 \item Hamiltonian formulation of duality symmetry by  Henneaux, Julia, Lekeu, and Ranjbar  \cite{Henneaux:2017kbx}, 2017, taking into account  different symplectic frames and different coset representatives,  following \cite{deWit:2002vt}.
  
\item The theoretical considerations are  combined in \cite{Kallosh:2023css} in 2023-2024 with the   amplitude computations in D-dimensional $\cN\geq 5$ supergravities  by Bern et al \cite{Bern:2023zkg}, 1998-2018.

\item RK et al  \cite{WIP1, WIP2}, very
recent work on the quantization of a local H-symmetry in symmetric and Iwasawa-type gauges  \cite{WIP1} and construction of the quantum path integral with manifest $Sp(2n_v, \mathbb{R}) $ symmetry \cite{ WIP2}.
This allowed the promotion of the proof of classical equivalence of various versions of  D-dimensional supergravities to the quantum level.

 \end{itemize}
 Thus, classical results in GZ duality, the existence of supergravities I and II, and the existence of different symplectic frames in 4D  related by GZ duality were applied to the quantization of supergravity in symmetric and Iwasawa-type gauges. The proof of gauge independence (absence of H-symmetry and G-symmetry anomalies) is possible in 4D $\cN\geq 5$ where duality is enhanced, dim [$Sp(2n_v)]\gg $ dim [$G_U$] and the double quotient  \cite{deWit:2002vt} in \rf{dq} is non-trivial. 

In $D> 4$,  the maximal duality is U-duality, for example in 6D $SO(5,5)\sim E_{5(5)}$.
Dualities are not enhanced, and the relevant double quotients of DeWit, Samtleben, and Trigiante are trivial. No proof is available of the on-shell gauge equivalence of symmetric and Iwasawa-type gauges and supergravities I and II. This implies a possibility of local H- global G-anomalies consistent with amplitude computations for maximal $D>4$  supergravities.  The evidence for these anomalies due to UV divergences was already presented in \cite{Kallosh:2023css}.

In conclusion, the existence of enhanced dualities in $N \geq 5$ supergravities in 4D provides a possible explanation of presently known cancellations of UV divergences in  4D supergravity.   These include $\cN=8$, L=3,4,5, which are usually explained by soft scalar limit due to \E\, symmetry as well as the enhanced case of $N = 5$, L=4, not explained by soft scalar limit due to \E\, symmetry \cite{Beisert:2010jx}. Thus, enhanced dualities act behind the scenes
 in amplitude computations in all 4D $N \geq 5$ cases.

\noindent Based on the current analysis, we suggest that:

1. New loop computations in D$>$4 supergravities may not bring new information: there is strong evidence of a  H-  and  G-symmetries anomalies, which makes perturbative D$>$4 supergravities inconsistent. 

2. New loop computations in 4D  $\cN\geq 5$ are highly desirable: if enhanced duality $Sp(2n_v)\supset G_U$  explains the absence of 4D UV divergences so far, 
we will see UV finiteness at the next loop orders.

\noindent{\bf {Acknowledgments:}} We are grateful to  J.J. Carrasco, D. Freedman, M. Gunaydin,  Y.-t. Huang,  A. Linde, R. Minasian, H.~Nicolai, R. Roiban, H. Samtleben, T. Van Proeyen, C. Wen and  Y. Yamada for stimulating discussions of  UV divergences and anomalies in supergravities.  This work is supported by SITP and by the US National Science Foundation grant PHY-2310429.

\end{document}